\documentstyle[12pt,aaspp4]{article}

\input epsf
\def\pref#1{(\ref{#1})}
\def\ep{\epsilon}
\def\ren{R_{m}}
\def\pref#1{(\ref{#1})}
\def\b{{\bf b}}
\def\ni{\noindent}
\def\lra#1{\left\langle #1\right\rangle}
\def\lrc#1{\left\{ #1\right\}}
\def\lrb#1{\left[ #1\right]}
\def\etal{{\it et al.}\ }
\def\bar{\overline}
\def\OV{\overline{\bf V}}
\def\OB{\overline{\bf B}}
\def\emf{\overline{\mbox{${\cal E}$}} {}}
\def\emfb{\overline{\mbox{\boldmath ${\cal E}$}} {}}
\def\bbE{\bar {\bf E}}
\def\bw{\bar{\omega}}
\def\beq{\begin{equation}}
\def\ee{\end{equation}}
\def\lsim{\mathrel{\rlap{\lower4pt\hbox{\hskip1pt$\sim$}}
    \raise1pt\hbox{$<$}}}
\def\gsim{\mathrel{\rlap{\lower4pt\hbox{\hskip1pt$\sim$}}
    \raise1pt\hbox{$>$}}}
\def\bfE{{\bf E}}
\def\bfJ{{\bf J}}
\def\bfH{{\bf H}}
\def\bfA{{\bf A}}
\def\bfa{{\bf a}}
\def\bfe{{\bf e}}
\def\bfB{{\bf B}}
\def\bbJ{\bar {\bf J}}
\def\bE{\bar E}
\def\bv{\bar V}
\def\bB{\overline B}
\def\ts{\times}
\def\lb{\langle}
\def\rb{\rangle}
\def\curl{\nabla {\ts}}
\def\nt{\nabla\times}
\def\cnt{\cdot\nt}
\def\Epsilon{\varepsilon}
\def\cd{\cdot}
\def\bbV{\bar {\bf V}}
\def\bfv{{\bf v}}
\def\bfvp{{\bf v}}
\def\bfV{{\bf V}}
\def\bfj{{\bf j}}
\def\bfjp{{\bf j}}
\def\bfe{{\bf e}}
\def\bfw{{\bomega}}
\def\bfwp{{\bomega}}
\def\bbw{\bar{\bf \bomega}}
\def\bfb{{\bf b}}
\def\bfbp{{\bf b}}
\def\bfB{{\bf B}}
\def\bfbv{\bf {\bar v}}
\def\bfbb{\bf {\bar b}}
\def\bbig{\bar{\bf \Omega}}
\def\bbB{\overline {\bf B}}
\def\bbA{\overline {\bf A}}

\def\div{\nabla\cdot}

\newcommand\figstages      {1}
\newcommand\figfinalstate  {2}
\newcommand\figa           {3}
\newcommand\figaa          {4}
\newcommand\fighs          {5}
\newcommand\figcch         {6}
\newcommand\figb           {7}
\newcommand\figc           {8}
\newcommand\fighv          {9}
\newcommand\fighb         {10}
\newcommand\figha         {11}

\newcommand\tableism      {1}
\newcommand\tablegalaxy   {2}
\newcommand\tablesim      {3}
\newcommand\tablescales   {4}

\newcommand\aaa{{\bf a}}
\newcommand\B{{\bf B}}
\newcommand\bb{{\bf b}}
\newcommand\f{{\bf f}}
\newcommand\kk{{\bf k}}
\newcommand\s{{\bf s}}
\newcommand\U{{\bf U}}
\newcommand\vv{{\bf v}}
\newcommand\x{{\bf x}}
\newcommand\zze{{\bf \zeta}}
\newcommand\bnabla{\mbox{\boldmath $\nabla$}}
\newcommand\hh{\hspace{1mm}}

\begin{document}

\setcounter{equation}{0}

\centerline{\large\bf A New Dynamical Mean-Field Dynamo Theory and 
Closure Approach}

\medskip

\author{Eric G. Blackman \altaffilmark{1} and George B. Field\altaffilmark{2}}
\affil{1. Department of Physics \&Astronomy and Laboratory for
Laser Energetics, University of Rochester, Rochester NY 14627}
\affil{2. Harvard-Smithsonian Center for Astrophysics, 60 Garden St.
Cambrdige MA, 02138}

\centerline {(in press, Phys. Rev. Lett.)} 

\begin{abstract} 

\noindent We develop a new nonlinear mean field dynamo theory 
that couples field growth to the time evolution of the magnetic helicity
and the turbulent electromotive force, $\emfb$.
We show that the difference between kinetic and current helicities 
emerges naturally as the growth driver when the time derivative of $\emfb$ is 
coupled into the theory. The solutions predict 
significant field growth in a kinematic phase and a   
saturation rate/strength that is magnetic Reynolds number 
dependent/independent in agreement with numerical simulations.
The amplitude of early time oscillations provides a diagnostic for the closure.

\end{abstract}

\centerline{PACS codes: 95.30.Qd;  98.38.Am; 52.55.Ip,
52.30.Cv; 98.35.Eg; 
96.60.Hv}
\bigskip



\ni {\it Introduction-}
Mean field dynamo (MFD) theory has been a useful  framework for modeling
the in situ origin of large-scale magnetic field 
growth in planets, stars, and galaxies 
[1-4], and has also been invoked 
to explain the sustenance of fields in  fusion devices
[\cite{ji,bellan}].  
However, whether the backreaction of the magnetic field itself prematurely
quenches the MFD has been debated [7-24].  Recent progress has emerged from 
incorporating  magnetic helicity evolution into the theory.



To make explicit the problem to be solved, we first average the magnetic 
induction equation to obtain the basic MFD equation
[\cite{moffatt,krause}]:
\beq
\partial_t\OB= \curl \emfb +\curl(\bbV\ts \bbB) -\lambda \nabla^2\OB,
\label{1}
\ee
where $\OB$ is the mean (large-scale) magnetic field in Alfv\'en speed units,
$ 
\lambda = {\eta c^2\over 4\pi}
$
is the magnetic diffusivity in terms of the resistivity $\eta$, $\bbV$ is
the mean velocity which we set $=0$, and 
$\emfb=\lb\bfv\ts\bfb\rb$
is the turbulent electromotive force, a correlation between
fluctuating velocity $\bfv$ and magnetic field $\bfb$ in Alfv\'en units.  
Textbook treatments [\cite{moffatt,krause}] invoke 
$\emfb=\alpha {\bbB}-\beta \curl \bbB$, 
where $\alpha$ and $\beta$ are pseudoscalar and scalar
correlations of turbulent quantities respectively. 
In the kinematic theory [\cite{moffatt}] $\alpha = -(\tau_c/3)\lb\bfv\cdot\curl\bfv\rb$, where $\tau_c$ is a correlation time, 
and (\ref{1}) is solved with $\alpha$ and $\beta$ as input parameters.

But the kinematic theory is incomplete.
In a study of
helical MHD turbulence, 
Ref. \cite{pfl} 
derived approximate evolution equations for the spectra of kinetic
energy, magnetic energy, kinetic helicity, and magnetic helicity
($\equiv \lb{\bfA\cdot\curl\bfA}\rb$, where  $\bfB=\curl \bfA$).  
These calculations suggested that 
$\alpha \simeq (\tau_c/3)\lb\bfb\cdot\curl\bfb-\bfv\cdot\curl\bfv\rb$,
the residual helicity, where $\lb \rb$ indicates spatial or ensemble average.
This form has been employed in attempts to understand  
nonlinear dynamo quenching by coupling  magnetic helicity conservation
into the dynamo through the $\lb\bfb\cdot\curl\bfb\rb$ term 
[\cite{zeldovich83},19-23]
Although these studies [e.g. \cite{zeldovich83,kleeorin82,kleeorinruz}],
wrote down an equation for the time evolution of $\alpha$,  
they derived quenching  formulae for $\alpha$  
only for the steady-state.  Only after a coupled nonlinear system of 
time-dependent large and small scale magnetic helicity 
equations were solved [\cite{fb02}], was it apparent that 
a dynamical quenching model based on residual helicity 
reveals both a kinematic growth phase and an asymptotic resistively limited 
phase as seen in numerical experiments [\cite{b2001}]. The dynamical approach
has also been applied to dynamos with shear [\cite{bb02}].

But even in these dynamical approaches,  
the $\emfb$ was assumed to be proportional to the residual helicity.
Here we show that the required residual helicity emerges 
not from $\emfb$, but from $\partial_t\emfb$, and that including the 
$\partial_t\emfb$ equation in addition to the MFD and total
magnetic helicity equations is essential for a complete MFD theory.
We first derive $\partial_t\emfb$ and then derive the triplet of 
equations to be solved for the simple
shear free helical dynamo whose solutions can be compared
with existing numerical simulations. We discuss these solutions, physical
implications, and the relation to previous work.



{\it Deriving $\partial_t \emfb$-}
A two-scale nonlinear quenching approach 
invoking the residual helicity in $\emfb$  [\cite{fb02}], 
captures the nonlinear dynamo saturation seen in 
simulations [\cite{b2001}], but 
the derivation of the appropriate $\emfb$  
has been elusive.  To couple  $\lb\bfb\cdot\curl\bfb\rb$ 
of $\emfb$ to the magnetic helicity conservation equation,
the full small-scale field $\bfb$ must enter this 
correlation not a low order approximation. 
The essence of the puzzle [\cite{paris01}] 
is that 
\beq
\emfb(t)=
\lb\bfv(t)\ts\bfb(t)\rb = \int_0^t \lb \bfv(t )\ts 
\partial_{t'}\bfb(t')\rb dt'= -\int_0^t \lb \bfb(t)\ts \partial_{t'}\bfv(t')\rb dt',
\label{c1}
\ee
assuming that $b(0)=0$ and that $t>>0$ so that 
$\lb\bfv(0)\ts\bfb(t)\rb=0$.
The last two terms 
are both exact expressions for  $\emfb$,
however neither leads naturally the residual 
helicity entering $\alpha$: upon using the 
induction equation for $\bfb$, 
the second term on the right leads to a term $\propto
\int\lb\bfv(t)\curl\bfv(t)\rb dt'$. Using the Navier-Stokes 
equation for $\bfv$, the last term contributes a term $\propto
\int\lb\bfb(t)\curl\bfb(t)\rb dt'$.
One emerges with a $choice$
rather than the $difference$ between the two helicities. 
So how does the residual helicity emerge?


Rather than impose the form of the $\emfb$, we solve for it dynamically 
using
\beq
\partial_t\emfb=\lb\partial_t\bfv\ts\bfb\rb +\lb\bfv\ts\partial_t\bfb\rb.
\label{timed}
\ee
To proceed, we need equations for $\partial_t\bfb$ and $\partial_t\bfv$.  
Assuming $\div\bfv=0$ we have
\begin{equation}
\begin{array}{r}\partial_{t} \bfb = \OB\cdot\nabla\bfv - \bfv\cdot\nabla\OB + 
\curl(\bfv\ts\bfb) 
-\curl\lb\bfv\ts\bfb\rb +
\lambda\nabla^{2}\bfb,
\end{array}
\label{c3}
\ee
and
\begin{equation}
\begin{array}{r}
\partial_{t} {v}_q=P_{qi}({\OB}\cdot\nabla{b_i} + {\bfb}\cdot\nabla{\bB_i} 
-\bfv\cdot\nabla v_i+\lb\bfv\cdot\nabla v_i\rb
+{\bf b}\cdot\nabla{b}_i-\lb {\bf b}\cdot\nabla{b_i}\rb)
+ \nu\nabla^{2}{v_q} 
+{f_q},
\end{array}
\label{c2}
\end{equation}
where $\bf f$ is a divergence-free 
forcing function uncorrelated with $\bfb$, $\nu$ is the viscosity, 
and $P_{qi}\equiv (\delta_{qi}-\nabla^{-2}\nabla_q\nabla_i)$
is the projection operator that arises after taking the divergence
of the incompressible Navier-Stokes equation to eliminate the total
fluctuating pressure (magnetic + thermal).
Using Reynolds rules [\cite{rad}] to interchange  brackets with time
and spatial derivatives, the 5th term of 
(\ref{c3}) and the 4th and 6th terms in the parentheses of 
(\ref{c2}) do not contribute when put into the averages so we 
ignore them.   

The contribution to $\partial_t\emfb$ 
from the 3rd term in (\ref{timed}) can be derived 
by direct use of (\ref{c3}) in configuration space.
We assume isotropy of the resulting velocity
and magnetic field correlations for terms linear in $\bbB$. We  also 
retain the triple correlations.  The contribution 
to $\partial_t\emfb$ from the 2nd
term in (\ref{timed}) also contributes terms linear in $\bbB$, 
and triple correlations.  Here the terms linear in $\bbB$ are best
derived in Fourier space. For this, 
we follow the technique in the appendix of Ref. \cite{gd2}, 
which invokes the Fourier transform of the terms linear in $\bbB$
 contributing to $\lb\partial_t\bfv\ts\bfb\rb$,  
supplemented by a linear expansion of the projection operator in $k_1<<k_2$, 
where $k_1$ is the characteristic wavenumber of the 
bracketed or mean quantities and 
$k_2$ is the characteristic wavenumber of the fluctuating quantities
$\bfb$ and $\bfv$. Collecting all surviving 
terms, we then have for (\ref{timed})
\beq
\begin{array}{r}
\partial_t{\emfb}=
{1\over 3}(\lb\bfb\cdot\curl\bfb\rb
-\lb\bfv\cdot\curl\bfv\rb)\bbB 
- {1\over 3}\lb
\bfv^2\rb\curl\bbB
+\nu\lb\nabla^2\bfv\ts\bfb\rb +\lambda\lb\bfv\ts\nabla^2\bfb\rb
+ {\bf T}^V+{\bf T}^M,
\end{array}
\label{timed33}
\ee
where ${{\bf T}^M}=\lb \bfv\ts\curl(\bfv\ts\bfb)\rb$ 
and ${T_{j}^V}=\lb(\ep_{jqn}\lb P_{qi}(\bfb\cdot\nabla b_i
-\bfv\cdot\nabla v_i) b_n\rb$
are  the triple correlations. Note that 
the 3rd, 4th, 6th and 8th  
terms in (\ref{timed33})
 come from  
the $\lb\bfv\ts\partial_t\bfb\rb$ term of (\ref{timed})
and the 2nd 5th and 7th terms come from the 
$\lb\partial_t\bfv\ts\bfb\rb$ term of (\ref{timed}).



We are primarily interested in the component of $\emfb$ parallel to 
$\bbB$. For this we have
\beq
\partial_t\emf_{||}=(\lb\partial_t\bfv\ts\bfb\rb
 +\lb\bfv\ts\partial_t\bfb\rb)\cdot\bbB/|\bbB|+
\lb\bfv\ts\bfb\rb\cdot\partial_t(\bbB/|\bbB|).
\label{timedp}
\ee
Substituting (\ref{timed33}) into 
(\ref{timedp}) gives 
\beq
\partial_t\emf_{||}= {\tilde\alpha}{\bbB^2/|\bbB|}
-{\tilde\beta}{\bbB\cdot\curl\bbB}/|\bbB|-{\tilde \zeta}\emf_{||}
\label{2}
\ee
where 
${\tilde\alpha}
=(1/3)(\lb\bfb\cdot\curl\bfb\rb-\lb\bfv\cdot\curl\bfv\rb)$,  ${\tilde\beta} = (1/3)\lb \bfv^2\rb$, and  
$\tilde \zeta$ 
accounts for microphysical dissipation
terms, the last term of (\ref{timedp}), and
any additional contribution  arising from
${\bf T}^M + {\bf T}^V\ne 0$. 
Note that $\tilde \alpha$ and $\tilde\beta$ appear similar
to the usual $\alpha$ and $\beta$ dynamo coefficients in $\emfb$, 
but they are fundamentally different because they
are coefficients in $\partial_t\emfb$ (and thus 
have different units) and do not involve $\tau_c$.
Note also that if isotropy of like correlations were strongly violated, 
$\tilde\alpha$ and $\tilde\beta$
would be anisotropic tensors in analogy to tensor generalizations 
of $\alpha$ and $\beta$ [\cite{kr,rk}]. We have not considered
that here.  

Following [\cite{fb02}] we 
define large and small-scale magnetic
helicities as $H_1^M\equiv \lb\bbA\cdot\bbB\rb_{vol}$
and $H_2^M\equiv\lb\bfa\cdot\bfb\rb_{vol}$,
where $\lb\rb_{vol}$ indicates a global spatial average.
Then 
$\lb\bfb\cdot\curl\bfb\rb_{vol} = k_2^2H_2^M$ and 
$\lb\bbB\cdot\curl\bbB\rb_{vol} = k_1^2H_1^M$.
We define the small-scale kinetic helicity
$H_2^V=\lb\bfv\cdot\curl\bfv\rb$.
We assume $\bbV=0$, and a force-free large-scale field 
for which the $\partial_tH_1^M$ equation 
becomes degenerate [\cite{fb02}] with that of $\partial_t\bbB^2$. 
Then $|\bbB|=k_1^{1/2}|H_1^{1/2}|$.
We can thus rewrite (\ref{timed}) as
\beq
\partial_t\emf_{||}=
k_1^{1/2}|H_1^M|^{1/2}(k_2^2H_2^M-H_2^V)/3-k_1^{3/2}(H_1^M/(|H_1^M|^{1/2})
{\tilde\chi}-{\tilde \zeta}\emf_{||}.
\label{4time}
\ee



{\it Dynamo Equations-}
We couple (\ref{4time})
to the equations for small and large-scale magnetic helicity 
evolution for a dynamo in which the kinetic energy is
externally forced and $\bbV=0$. 
We interpret $\bbB$, $\bbA$ and $\emfb$ 
as the $k_1$ ($0 < k_1 < k_2$) component of $\bfB, \bfA$ and $(\bfv\ts\bfb)$ 
of a closed system to facilitate  comparison with simulations of Ref. 12.  The total magnetic helicity, $H^M=\lb\bfA\cdot\bfB\rb_{vol}$, 
then satisfies [\cite{moffatt}]
$
\partial_t\lb\bfA\cdot\bfB\rb_{vol}=-2\lb{\bfE}\cdot\bfB\rb_{vol},
$
where $\bf E= -\partial_t {\bf A} -\nabla \phi$,
and  $\phi$ is the scalar potential.
The large and small scale 
integrated magnetic helicity equations are then [\cite{b2001,fb02,bf00}]
\beq
\partial_t\lb\bbA\cdot \bbB\rb_{vol}=2\lb{\emfb}\cdot\bbB\rb_{vol}-2\lambda \lb{\bbB\cdot \curl\bbB}\rb_{vol},
\label{h1}
\ee
and
\beq
\partial_t\lb\bfa\cdot \bfb\rb_{vol}=-2\lb{\emfb}\cdot\bbB\rb_{vol} -2\lambda 
\lb{\bfb\cdot\curl\bfb}\rb_{vol}.
\label{h2}
\ee
When $\bbB$ is force-free, the two-scale
approximation allows us to write 
(\ref{h1}) and (\ref{h2}) as 
\beq
\partial_t H_1^M=2\emf_{||}k_1^{1/2}|H_1^M|^{1/2}
-2\lambda k_1^2H_1^M
\label{h3}
\ee
and 
\beq
\partial_t H_2^M=-2\emf_{||}k_1^{1/2}|H_1^M|^{1/2}-2\lambda k_2^2H_2^M.
\label{h4}
\ee
We need to solve (\ref{h3}), (\ref{h4}), and (\ref{4time})
after  converting them into dimensionless form.
We define the dimensionless quantities
$h_1\equiv H_1^M (k_2/v_2^2)$ and $h_2\equiv H_2^M (k_2/v_2^2)$,
$\ren\equiv v_2/\lambda k_2$, $Pr_M\equiv \nu/\lambda$, 
$\tau\equiv tv_2k_2$, $Q=-\emf_{||}/v_2^2$, 
$\chi={\tilde\beta}/v_2^2$, $\zeta = {\tilde \zeta}/v_2 k_2=(1+Pr_M)/\ren$, and 
use $H_2^V=-k_2v_2^2$.
For (\ref{h3}), (\ref{h4}) and (\ref{4time}) respectively, 
this gives
\beq
\partial_\tau h_1=-2Qh_1^{1/2}(k_1/k_2)^{1/2}-2h_1(k_1/k_2)^2/\ren,
\label{7}
\ee
\beq
\partial_\tau h_2=2Qh_1^{1/2}(k_1/k_2)^{1/2}-2h_2/\ren,
\label{8}
\ee
and
\beq
\partial_\tau Q= -\left(k_1/k_2 \right)^{1/2}h_1^{1/2}(1+h_2)/3
+(k_1/k_2)^{3/2}h_1^{1/2}\chi-\zeta Q.
\label{6}
\ee


{\it Solutions-}
Since $H_2^V < 0$, $H_1^M>0$ and $H_2^M<0$ for
a growing solution.  The solutions of (\ref{7}), (\ref{8}), and (\ref{6}) for
two different $\ren$ are shown
in Figs. 1 \& 2 over different time ranges 
for both $\zeta=2/R_M$ and $\zeta=1$
we use ${\tilde \alpha}\propto {\tilde \beta}$ in Fig. 1,
but the resulting solutions are only weakly sensitive to 
the form of $\tilde \beta$ as shown in Fig. 2.
In Figs.1\&2 we also plotted the empirical fit formula
to numerical simulations [\cite{b2001}]
(equation (54) of Ref \cite{b2001}) using our dimensionless parameters,
to demonstrate the good agreement.

In Figs. 1\&2 we  also compare the triplet
solution of $h_1$ with the doublet solution [\cite{fb02}]
that results from solving (\ref{7}) and (\ref{8}) but 
with $imposing$ $\emfb=\alpha\bbB-\beta\curl\bbB$ such that  
$Q=Q_{d}\equiv -(k_1/k_2)^{1/2}h_1^{1/2}(1+h_2)\tau_c/3+
(k_1/k_2)^{3/2}h_1^{1/2}\chi\tau_c$,
%
%
where the correlation time $\tau_c$ is a
free parameter taken to be $\sim 1$.
Note that the present triplet solution does not involve
$\tau_c$ in the dynamo coefficients $\tilde\alpha$ and $\tilde\beta$.
But a remarkable result emerges: 
Fig. 1 shows that the triplet solution  matches the doublet solution
at early times for $\zeta=1$, which corresponds to 
a damping time $\zeta^{-1}\sim \tau_c$. This arises from a closure
in which the triple correlations ${\bf T}^M$ and  ${\bf T}^V$
lead to a damping with time constant $\sim \tau_c$,
and the damping suppresses the oscillations.
Fig. 2 also shows that 
the  $\zeta=2/\ren$ and $\zeta=1$ cases are indistinguishable at late times.
We can also compare the kinematic regimes of the triplet and
doublet solutions.
The rise to the first peak of $h_1$ in Fig. 1a
is independent of $\ren$ and there the two $\ren$ triplet solutions
overlap.  This is the  kinematic regime.  
The end of the doublet kinematic regime
occurs at $h_1\simeq 1$, as
seen in Fig. 1.  Again, the doublet and triplet match when $\zeta\sim 1$.
In sum:
{\it The doublet solution emerges as the limit of the triplet
solution when the triple correlations act as a damping term.}
This closure can be tested with future simulations.


The maximum kinematic growth rate for $h_1$ is a function of $\zeta$
because it occurs where 
$Q$ is a minimum. 
If we ignore resistive terms so that $h_1 = -h_2$,
and assume that $\zeta<<1$ and $\chi=1/3$, 
then Eq. (\ref{6}) implies that 
the maximum growth rate occurs when $h_1\simeq 1 - k_1/k_2$.  From Fig 1a, 
the minimum of $Q$ during the first oscillation is 
$\sim -1/3$ (found to be independent of $k_1/k_2$).  Setting 
$\partial_\tau h_1 \sim n h_1$, the maximum kinematic growth rate  from
(\ref{7}) is then $n \sim  (2/3)(k_1/k_2)^{1/2}(1 - k_1/k_2)^{-1/2}
\sim 0.33$, for $k_2=5k_1$.
However, when $\zeta=1$, the minimum of $Q$ from (\ref{6}) 
occurs where $Q=Q_d$. In this case, $n\sim (2/3)(k_1/k_2)(1-k_1/k_2)\sim 0.11$
for $k_2=5k_1$. 
This demonstrates that the kinematic growth rate for
$\zeta<<1$ is $\sim 3$ times that for $\zeta=1$, and 
why the triplet kinematic growth rate with $\zeta=1$ matches that of the 
doublet. Both results are seen in Fig 1.

Inspection of (\ref{7}), (\ref{8}), and (\ref{6}) 
reveals why there are oscillations for a positive seed $h_1$
and $\zeta<<1$ (and independent of whether $\emfb(t=0)=0$ or $\emfb(t=0)\ne 0$). 
As long as $-1< h_2 <0$, $Q$ grows more negative and 
$h_1$ and  $h_2$  grow with mutually opposite signs.
As $h_2$ passes through $-1$ from above,   
$\partial_\tau Q$  changes sign immediately but 
$h_1$ continues to grow positive, albeit more slowly, 
until $Q$ changes sign.  Then, $\partial_\tau h_1$ 
changes sign and $h_1$ decreases. But $\partial_\tau h_2$  
changes sign when $\partial_\tau h_1$  does, 
so when $h_2$ eventually passes back through $-1$
from below,  $\partial_\tau Q$ reverses sign again, and 
eventually $Q$ becomes negative and $h_1$ again grows.  
Large $\ren$ terms only weakly damp the oscillations.
This describes what happens in Fig 1a.   
If instead, $\zeta \sim 1$, 
once $\partial_\tau Q$ is depleted 
by the growth of $h_2$, the $\zeta$ term of
(\ref{6}) takes over and $Q$ decays without oscillating.  
Then $h_1$ grows without oscillations (Fig 1b).



{\it Discussion-}
Textbook kinematic MFD theories solve only
the MFD equation itself [\cite{moffatt}]. 
Recent nonlinear approaches incorporating magnetic helicity evolution 
dynamically [\cite{fb02,bb02}] solve a doublet: 
the MFD equation (or the $\partial_t H_1^M$ equation)
and the total magnetic helicity evolution equation
(the $\partial_t H_1^M + \partial_t H_2^M$ equation).  
The present paper solves a triplet: the MFD equation, the total 
magnetic helicity evolution equation, $and$ the $\partial_t \emfb$ equation. 
Only the present approach shows how the difference between 
kinetic and current helicities (the residual helicity)
emerges as the MFD driver in a time dependent theory.  
The residual helicity in turn couples to the total  
magnetic helicity evolution equation.
The physical interpretation of the solutions for a closed  system 
is that as the large scale
helical field grows from MFD action, the small scale magnetic helicity 
grows of the opposite sign. At early times, kinematic growth is 
unimpeded, and the large scale field energy grows to 
$\bB^2 \gsim (k_1/k_2)v_2^2$.  Eventually, the 
small-scale magnetic helicity backreacts on the kinetic
helicity, suppressing the growth rate to an $R_M$ dependent
value. Ultimately $\bB^2 \simeq (k_2/k_1)v_2^2 $ at saturation.
This picture also arises in the imposed $\emfb$  doublet approach
[\cite{fb02}]. The approaches agree in the  
asymptotic $R_m$ dependent growth phase, matching 
simulations [\cite{b2001}]. However, oscillations  
are possible at early times only in the triplet approach.
The amplitude of such oscillations 
serves as a direct diagnostic for the MHD closure scheme, which can be 
tested with future numerical experiments.
For $\zeta=1$, the agreement between the two approaches becomes
exact for all times. This  corresponds to the triple correlations
contributing a simple damping term with characteristic time scale 
$\sim \tau_c$.

\ni 
We thank A. Brandenburg for discussions and for
pointing out a proper derivation of (\ref{timed33}).
EB acknowledges DOE grant DE-FG02-00ER54600, 
and the Aspen Center for Physics and NORDITA for hospitality.

\enumerate


\bibitem{moffatt} 
H.K. Moffatt, {\sl Magnetic
Field Generation in Electrically Conducting Fluids}, 
(Cambridge University Press, Cambridge, 1978)

\bibitem{parker}  
E.N. Parker, {\it Cosmical Magnetic Fields}, (Oxford: Clarendon
Press, 1979)

\bibitem{krause}   F. Krause \& K.-H. R\"adler, 
{\it Mean-field Magnetohydrodynamics and Dynamo Theory}, 
(Pergamon Press, New York, 1980)

\bibitem{zeldovich83} 
Ya. B. Zeldovich , A.A. Ruzmaikin, \& D.D. Sokoloff, {\sl Magnetic Fields in Astrophysics}, 
(Gordon and Breach, New York, 1983)

\bibitem{ji} 
H. Ji, Phys. Rev.  Lett., {\bf 83}, 3198 (1999);
H. Ji in {\it Magnetic Helicity in Space and Laboratory Plasmas}, 
edited by A. Pevtsov, R. Canfield, \& X. Brown, 
(Amer. Geophys. Union, Washington, 1999), p167; 
H. Ji, \& S.C. Prager, in press  Magnetohydrodynamics, astro-ph/0110352
(2002)

\bibitem{bellan} P.M. Bellan, {\sl Spheromaks}, 
(Imperical College Press, London, 2000)

\bibitem{piddington} 
   J.H. Piddington,  {\it
Cosmical Electrodynamics}, (Krieger Press, Malbar, 1981) 

\bibitem{vc}
S.I. Vainshtein  \& F. Cattaneo, ApJ, {\bf 393}, 165 (1992)

\bibitem{ch} 
 F. Cattaneo, \& D.W. Hughes, Phys. Rev. E., {\bf 54}, 4532 (1996)

\bibitem{cattaneo2002}
F. Cattaneo, D.W. Hughes, D.W. \& J.-C. Thelen, JFM, {\bf 456}, 219 (2002)

\bibitem{kulsrud} 
   R.M. Kulsrud \& S.W. Anderson, ApJ, {\bf 396}, 606 (1992)

\bibitem{b2001}
A. Brandenburg, ApJ, {\bf 550}, 824 (2001)

\bibitem{fbc}
G.B. Field, E.G. Blackman \& H. Chou, ApJ, {\bf 513}, 638 (1999)

\bibitem{bf2000}
E.G. Blackman and G.B. Field, ApJ, 
{\bf 534}, 984 (2000)

\bibitem{fb02} 
G.B. Field  \& E.G. Blackman, ApJ, {\bf 572}, 685 (2002)

\bibitem{bb02}E.G. Blackman \& A. Brandenburg,
in press, ApJ, (2002),
http://xxx.lanl.gov/abs/astro-ph/0204497

\bibitem{paris01}E.G. Blackman, To 
appear in {\it Turbulence and Magnetic Fields in Astrophysics}
eds.~E.~Falgarone and T.~Passot, Springer Lecture Notes in Physics (2002), 
http://xxx.lanl.gov/abs/astro-ph/0205002

\bibitem{pfl} 
 A. Pouquet, U. Frisch, \& J. Leorat,  JFM, {\bf 77}, 321 (1976)

\bibitem{kleeorin82}
N.I. Kleeorin  \& A.A. Ruzmaikin, 
{Magnetohydrodynamics}, {\bf 18}, {116}, (1982)

\bibitem{kleeorinruz}
N. Kleeorin I.  Rogachevskii, A. Ruzmaikin,
A\&A,  {\bf 297}, L59 (1995)

\bibitem{gd1}  
 A.V. Gruzinov \&  P.H. Diamond P.H., Phys. Rev. Lett., {\bf 72}, 1651 (1994);

\bibitem{gd2}  
A.V. Gruzinov \& P.H. Diamond, Phys. of Plasmas, {\bf 2}, 1941 (1995);

\bibitem{by}  A. Bhattacharjee \& Y. Yuan, ApJ,  
{\bf 449}, 739 (1995).

\bibitem{vishniac}  E. Vishniac \& J. Cho, ApJ, {\bf 550}, 752 (2000).

\bibitem{rad} K.-H. R\"adler, K.-H. Astron. Nachr., {\bf 301}, 101 (1980)

\bibitem{kr}
N. Kleeorin \& I. Rogachevskii,  Phys. Rev. E., {\bf 59}, 6724 (1999)

\bibitem{rk} I. Rogachevskii 
\& N. Kleeorin, Phys. Rev E., {\bf 64}, 56307 (2001)
K.-H. Raedler, N. Kleeorin, \& I. Rogachevskii, 
sub. to Geophys. Ap. Fluid Dyn. (2002).

\bibitem{bf00} E.G. Blackman,  \& G.B. Field, MNRAS, {\bf 318}, 
724 (2000)

\bigskip
\bigskip
\bigskip
\bigskip

\vspace{-.1cm} \hbox to \hsize{ \hfill \epsfxsize7.5cm
\epsffile{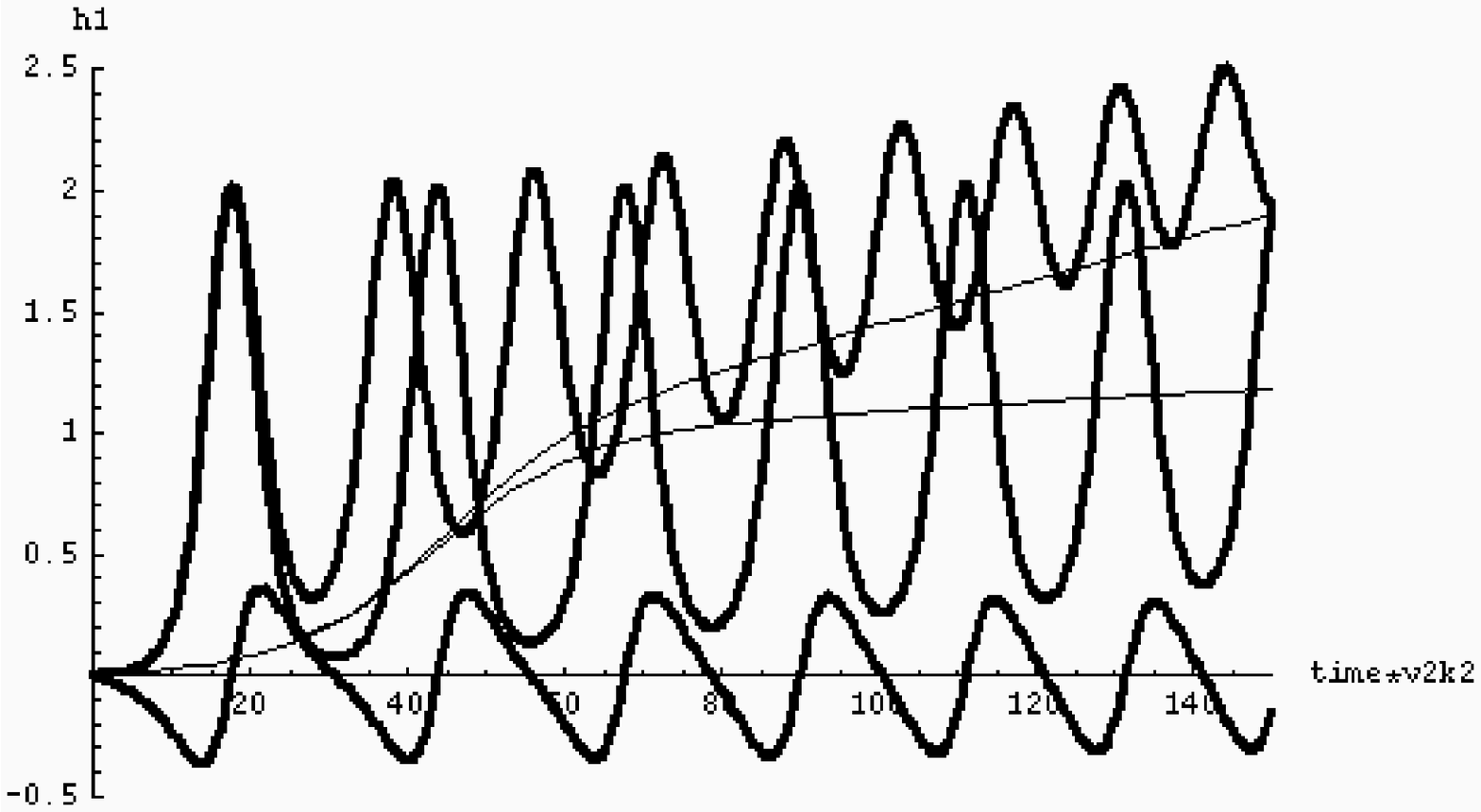} \epsfxsize8cm \epsffile{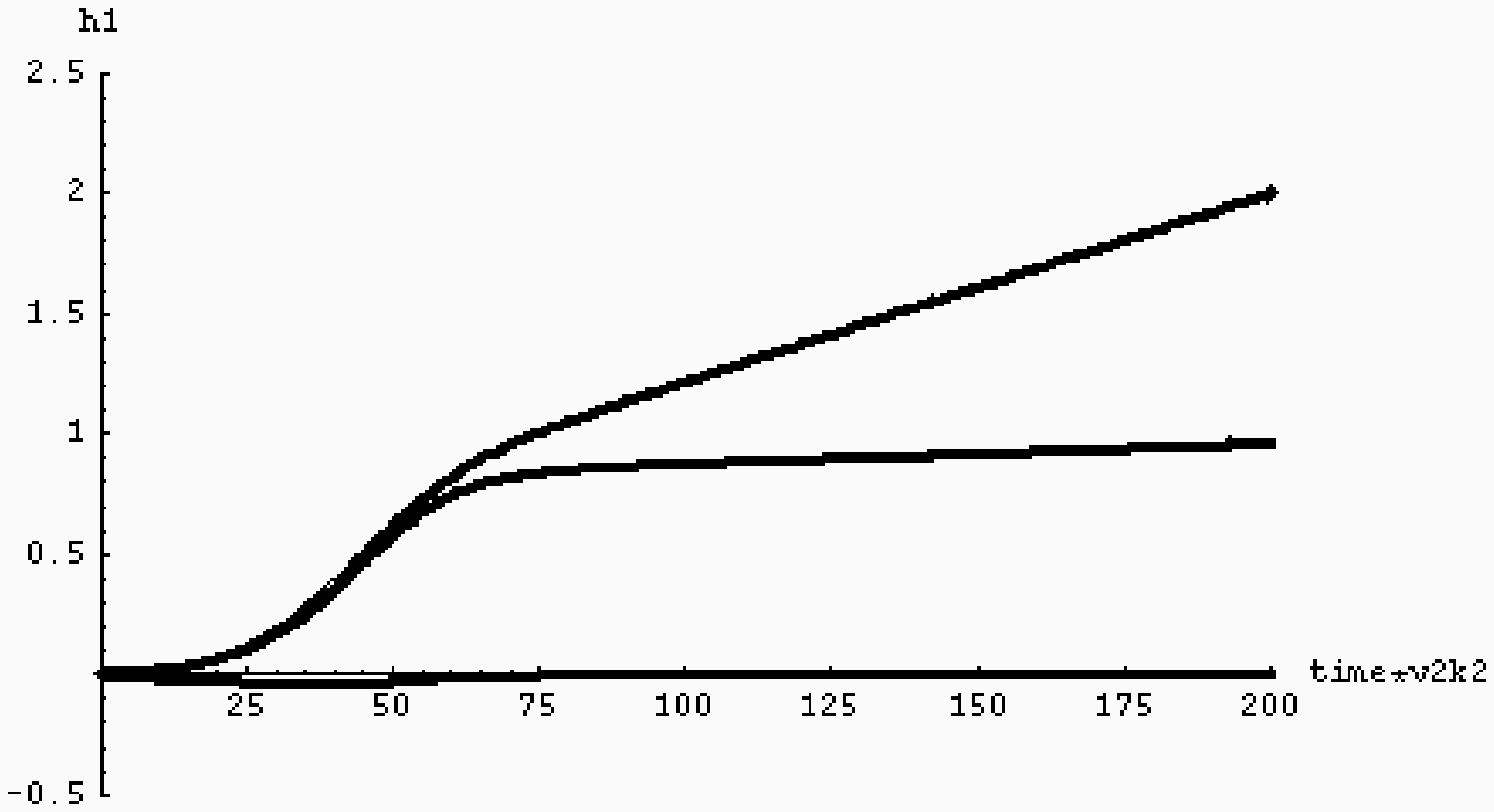} \hfill }  

\noindent {Figure 1: (a) Plot for $k_1=1$, $k_2=5$ with $\zeta=2/\ren$. 
The top and middle oscillating curves are for
 $h_1$ with $\ren = 200$  and $\ren = 1000$, and the 
bottom oscillating curve is  $Q$ for $\ren=1000$.
The thin lines are the doublet solutions 
for $h_1$ from Ref. \cite{fb02} which used an imposed $\emfb$. 
Here ${\chi} \propto (1+h_2)$.
(b) Same as (a) but with $\zeta\sim 1$.}
\vspace{-.1cm} \hbox to \hsize{ \hfill \epsfxsize8cm
\epsffile{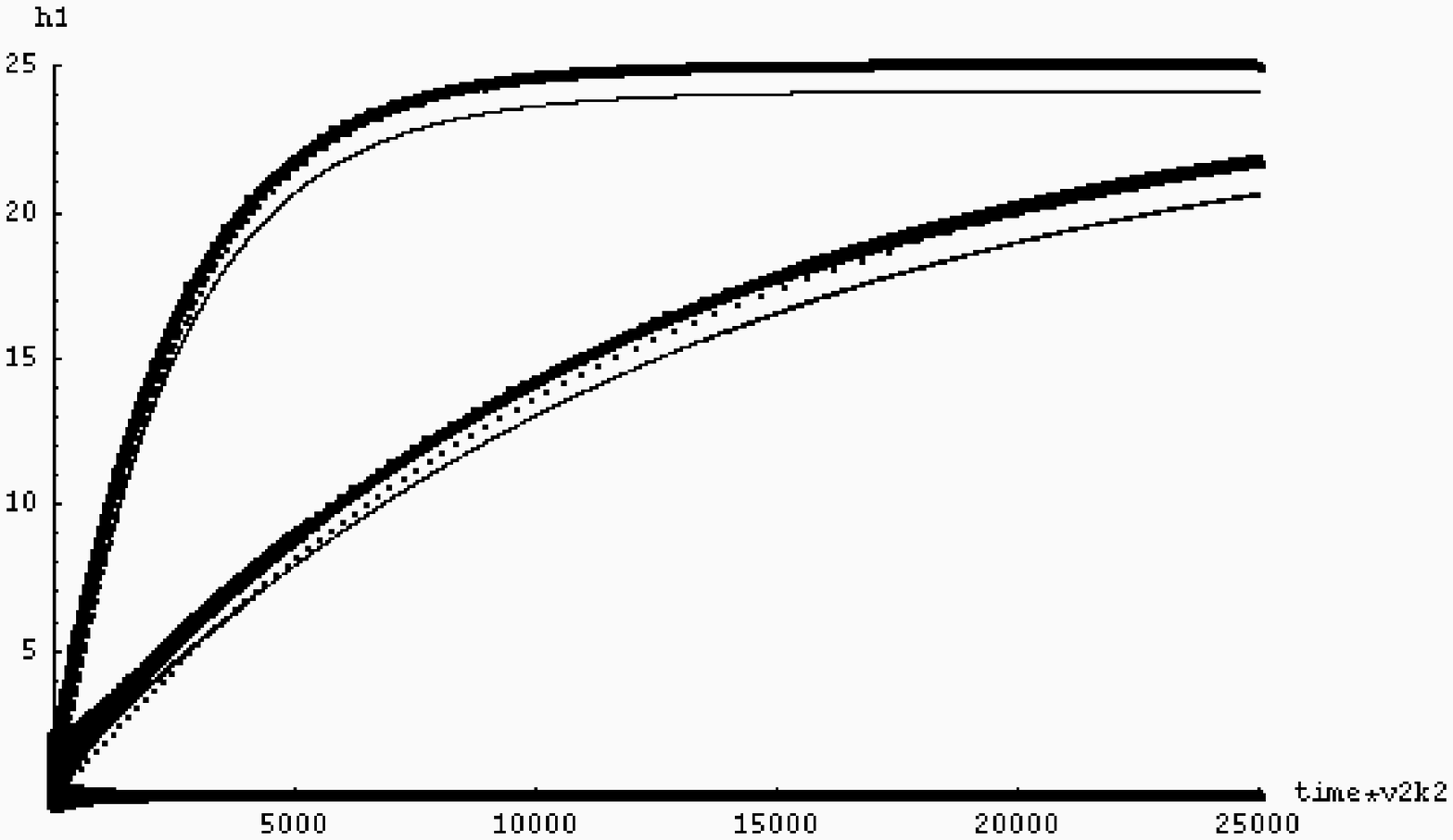} \epsfxsize8cm \epsffile{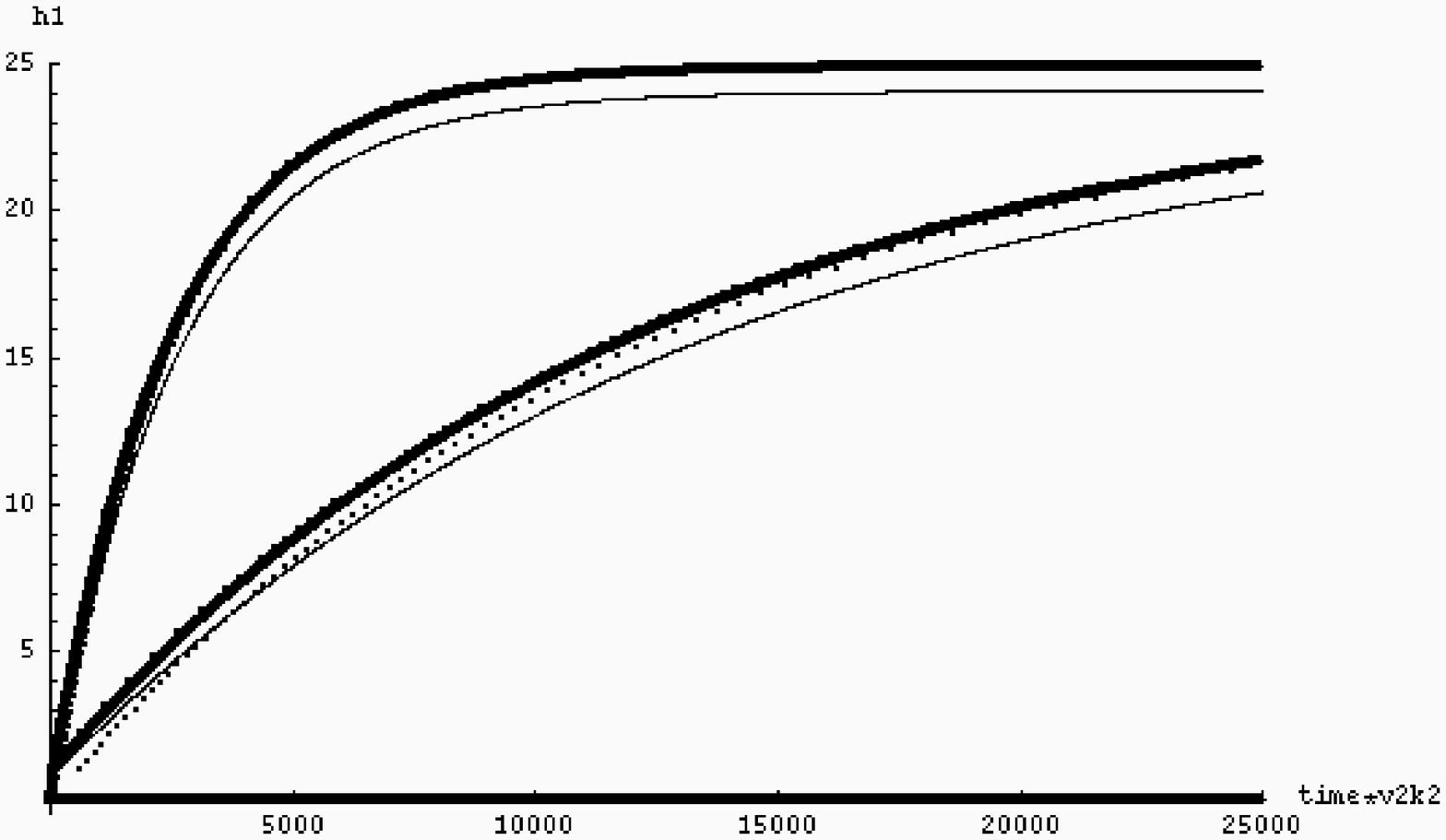} \hfill }

\noindent { Figure 2:
(a) Same as Fig. 1a but for broader time range.  
(b) Same as (a)  but for $\zeta=1$.  
For this time range, the doublet and triplet solutions are indistinguishable.
The dotted curves are fits to simulations [\cite{b2001}].
The lines slightly below each of the thick lines are for 
$\chi\propto 1/(1+k_1 h_1/k_2)$ demonstrating 
the weak dependence on $\tilde \beta$.


\end{document}